\documentclass[aps,pra,onecolumn,twoside,floatfix,a4,showpacs,superscriptaddress]{revtex4}

\expandafter\let\csname equation*\endcsname\relax
\expandafter\let\csname endequation*\endcsname\relax

\usepackage[dvips]{graphics}
\usepackage{amsmath,color}
\usepackage{amsfonts}
\usepackage{soul}

\usepackage{amssymb}
\usepackage{amstext}
\usepackage{amsthm}

\usepackage{braket}
\usepackage{bbold}
\usepackage{mathtools}
\usepackage{gensymb}
\usepackage{csquotes}

\newcommand{\tr}{\mathrm{tr}}
\newcommand{\ignore}[1]{}

\newcommand{\bs}{\bigskip}
\newtheorem{theorem}{Theorem}

\makeatletter
\newcommand*{\defeq}{\mathrel{\rlap{%
                     \raisebox{0.3ex}{$\m@th\cdot$}}%
                     \raisebox{-0.3ex}{$\m@th\cdot$}}%
                     =}
\makeatother
\begin{document}

\title{Smallest disentangling state spaces for general entangled bipartite quantum states}

\author{Hussain Anwar}
\affiliation{Department of Physics, Imperial College London, Prince Consort Road, London SW7 2AZ, UK.}
\author{Sania Jevtic}
\affiliation{Department of Mathematics, Huxley Building, Imperial College, London SW7 2AZ, UK.}
\affiliation{Institut f\" ur Theoretische Physik, Leibniz Universit\" at Hannover, Appelstra\ss e 2, 30167 Hannover, Germany.}
\author{Oliver Rudolph}
\affiliation{Leonardo da Vinci Gymnasium, Im Spitzerfeld 25, 69151 Neckargem\"und, Germany.
\& Hector-Seminar, Waldhoferstr.~100, 69123 Heidelberg, Germany.}
\author{Shashank Virmani}
\email{shashank.virmani@brunel.ac.uk}
\affiliation{Department of Mathematical Sciences, Brunel University, Uxbridge, Middlesex UB8 3PH, UK.}

\begin{abstract}
Entangled quantum states can be given a separable decomposition if we relax the restriction that the local operators be quantum states. Motivated by the construction of classical simulations and local hidden variable models, we construct `smallest' local sets of operators that achieve this. In other words, given an arbitrary bipartite quantum state we construct convex sets of local operators that allow for a separable decomposition, but that cannot be made smaller while continuing to do so. We then consider two further variants of the problem where the local state spaces are required to contain the local quantum states, and obtain solutions for a variety of cases including a region of pure states around the maximally entangled state. The methods involve calculating certain forms of cross norm. Two of the variants of the problem have a strong relationship to theorems on ensemble decompositions of positive operators, and our results thereby give those theorems an added interpretation. The results generalise  those obtained in our previous work on this topic [New J. Phys. \textbf{17}, 093047 (2015)].
\end{abstract}

\pacs{03.67.-a, 03.65.-w, 03.65.Ta, 03.65.Ud.}

\maketitle

\section{Background and Overview} \label{Background}

In recent years a variety of work has examined the nature of quantum entanglement by considering quantum systems as examples of more general structures. In the study of generalised probabilistic theories \cite{Boxworld}, for example, various axioms are postulated (such as no instant signalling) for reasonable physical theories, without necessarily demanding that the theory has an underlying Hilbert space or operator space structure. Other works have considered theories with an operator structure, but where the operators are allowed for instance to have negative eigenvalues \cite{OperatorGPT}, so long as the measurements are restricted in a way that leads to valid probabilities. The work of \cite{Barnum} was one of the first to consider the consequences of this line of thinking for entangled quantum states. A feature emerging from these previous constructions is the idea that if the available observables we have are restricted, then, in principle, our perception of the entanglement in a given system may change. In other articles this idea motivated us to consider whether this can help design new local hidden variable (LHV) models \cite{B64} and classical simulation methods for entangled quantum states measured in restricted ways \cite{Shash1,Shash2,AJRVpeps}.

One of the observations underlying the present work is that by allowing local state spaces to contain non-quantum operators, an entangled quantum state may in fact admit a separable decomposition that can be considered to be \textit{unentangled} if the measurements available are restricted. This motivates the question: if we want to use such generalised separable decompositions for the construction of LHV models or for efficient classical simulation algorithms, what such decompositions are useful? This question motivated a problem that we considered in a previous work \cite{AJRVsmall}: what are the {\it smallest} sets of operators such that a given entangled quantum state admits a separable decomposition? In this work we will generalise the results of \cite{AJRVsmall} to a much wider family of states.

\medskip

{\bf Structure}---This paper is structured as follows. In section \ref{sec_gen_sep} we explain and motivate the four problems (Problems 1,2,2*,3) that we consider in this work. In section \ref{ournorms} we present some norms that we use to characterise how small a given set of states is. In section \ref{results} we present a summary of our main results, which constitute three Theorems. Theorem \ref{th1} characterises all solutions to Problem 1 for all bipartite states for a subset of the norms that we consider, and it is proven in section \ref{characterisation}. Theorem \ref{th2} solves Problem 2 for all bipartite states, and is proven in section \ref{Prob2}. Theorem \ref{th3} presents solutions to problems 2* and 3 for certain families of bipartite state (including the pure states in a region around the maximally entangled state), and is proven in section \ref{prob3}. Those not interested in proofs may skip sections \ref{characterisation},\ref{Prob2}, and \ref{prob3}.

\section{Generalised Separability and Problems Considered} \label{sec_gen_sep}

In this work we will be concerned with a bipartite setting in which there are two systems $ A $ and $ B $ with arbitrary finite local Hilbert space dimensions. Their joint Hilbert space is $ \mathcal{H}_{AB}=\mathcal{H}_A\otimes\mathcal{H}_B $ and their quantum state is described by a density operator $ \mathcal{\rho}_{AB}$ with $\rho_{AB} \geq 0$ and $\tr(\rho_{AB})=1$. According to the standard definition of quantum entanglement, a state $ \rho_{AB} $ is said to be \textit{separable} if
\begin{eqnarray}
\rho_{AB} = \sum_j p_j \rho^A_j \otimes \rho^B_j,
\end{eqnarray}
where the local quantum operators $\rho^A_j$ and $\rho^B_j$ are positive semidefinite with $\tr(\rho^A_j)=\tr(\rho^B_j)=1$, and $p_j$ forms a probability distribution, otherwise it is said to be \textit{entangled} \cite{W89}. In this work we will be interested in generalising these definitions, and so we will instead use the expressions {\it quantum}-separable or {\it quantum}-entangled. The reason for this is that if one relaxes the requirement that $\rho^A_j$ and $\rho^B_j$ be positive operators, then one can indeed construct `separable' decompositions for any bipartite state $ \rho_{AB} $. An illustrative example can be constructed from the operator-Schmidt decomposition of any given quantum state $ \rho_{AB} $:
\begin{eqnarray}
\rho_{AB} =  \sum^D_{i=1} \lambda_i X_i \otimes Y_i,
\end{eqnarray}
where $X_i$ and $Y_i$ are local orthonormal operator bases (i.e. $\tr(X^{\dag}_i X_j)=\tr(Y^{\dag}_i Y_j)=\delta_{ij}$), the Schmidt coefficients $\lambda_i > 0$ are real positive numbers, and the operator-Schmidt rank $D$ satisfies $D \leq \dim (\mathcal{H}_A)^2,\dim (\mathcal{H}_B)^2 $. Defining $\lambda:=\sum_i \lambda_i$, and $\widetilde{X}_i := \sqrt{\lambda} X_i$, $\widetilde{Y}_i := \sqrt{\lambda} Y_i$, we can rewrite this decomposition in the separable form:
\begin{eqnarray}
\rho_{AB} =  \sum^D_{i=1} {\lambda_i \over \lambda} \widetilde{X}_i \otimes \widetilde{Y}_i. \label{example}
\end{eqnarray}
As the coefficients ${\lambda_i / \lambda}$ form a probability distribution, Eq. \eqref{example} is a separable decomposition in terms of typically non-positive operators $\widetilde{X}_i, \widetilde{Y}_j $. Although Eq. (\ref{example}) may seem like a trivial rewriting of the Schmidt decomposition, we shall show later that this decomposition has an interesting property:
\begin{displayquote}

If we consider the local operator sets formed from the convex hulls of the $\widetilde{X}_i$ and $\widetilde{Y}_i$ for each subsystem separately, then the resulting sets cannot be made \textit{smaller} while continuing to admit a separable decomposition for $\rho_{AB}$.
\end{displayquote}

Hence we say that the decomposition Eq. (\ref{example}) is an example of a `smallest' separable decomposition for the state $\rho_{AB}$. Finding such decompositions for various definitions of `smallest' will be the central aim of this article.

The motivation for looking for such smallest separable decompositions arises from their utility in the construction of LHV models or efficient classical simulation algorithms (see, for example, \cite{AJRVsmall,AJRVpeps}). The basic principle that underpins such constructions is the idea that a non-positive operator (a \textit{non-physical} quantum state---i.e. cannot be \textit{prepared} by a physical process) can in fact be considered a valid description of quantum system when the available measurements are restricted. Variations of this idea have a long history, going back at least as far as the Wigner function \cite{Wigner}. For convenience we briefly review why this is.

Consider an experiment in which a quantum state $\rho$ undergoes a physical process $ \mathcal{P} $ consisting of a transformation $L:\rho\mapsto L(\rho)$, followed by a measurement yielding an outcome corresponding to a measurement operator $M \geq 0$. The whole process, denoted by $\mathcal{P}(L,M)$, occurs with probability  $\tr(M L(\rho))/\tr(L(\rho))$. In such situations, it would be instructive to decompose $\rho$ into a probabilistic ensemble of positive density matrices $\rho_i$, each prepared with classical probability $p_i$, such that $\rho = \sum_i p_i \rho_i$. Due to linearity, one can analyse the experiment in terms of the process $\mathcal{P}(L,M)$ acting individually on each of the $\rho_i$ operators in the decomposition. This approach usually assumes that each $\rho_i$ is a positive operator, however, it works even when the operators $\rho_i$ that appear in the decomposition are \textit{not} positive, provided that the processes $\mathcal{P}(L,M)$ being considered are restricted so that all the $\rho_i$ satisfy $\tr(M L(\rho_i)) \geq 0$ and $\tr(L(\rho_i))> 0$. These properties ensure that the classical manipulation of the probabilities $p_i$ follows the usual Bayesian rules, and hence there are no obstacles to considering the system as a mixture of the otherwise non-physical operators $\rho_i$ \footnote{The existence of non-positive operators satisfying these conditions is possible when restricted $(L,M)$ are considered. In situations where the full set of quantum operations and measurements are considered, then the $\rho_i$ are necessarily positive.
}.

We refer to the properties $\tr(M L(\rho_i)) \geq 0$ and $\tr(L(\rho_i)) > 0$ as notions of {\it generalised} positivity with respect to $\mathcal{P}(L,M)$. In this paper the choice of $L$ and $ M $ will be either arbitrary or clear from the context, and so for brevity we will usually just use the term {\it generalised positivity} without explicitly mentioning the  transformations $L$ or measurements $M$.

\bs

\textit{LHV construction and motivating examples for Bell states}--- One of the motivations for looking for such generalised separable decompositions is that they can provide LHV models for measurements for which the local operators are positive. For convenience we now explain this standard argument (as used, for example in the use of Wigner functions to construct LHV models \cite{Bell}). Suppose that
\begin{equation}
\rho_{AB} = \sum_i q_i A_i \otimes B_i, \nonumber
\end{equation}
is a separable decomposition of a quantum state $\rho_{AB}$, where $A_i$ and $B_i$ are generalised-positive for local positive-operator valued measures (POVMs) $\mathcal M=\{M_i|M_i \geq 0, \,\sum_i M_i=\mathbb{1}\}$ and $\mathcal N=\{N_i|N_i \geq 0, \sum_i N_i=\mathbb{1}\}$. The $q_i$ must be positive but need not be normalised.
The probability of obtaining outcomes $M_k \in \mathcal M$ and $N_l \in \mathcal N$ is
\begin{eqnarray}
\tr (\rho_{AB} M_k \otimes N_l) &=& \sum_i q_i \tr (A_i M_k) \tr (B_i N_l), \nonumber \\
&=& \sum_i q_i a(k|i) b(l|i), \label{lhv}
\end{eqnarray}
where $a(k|i):= \tr (A_i M_k)\geq 0$ and $b(l|i):=\tr (B_i N_l)\geq 0$. The completeness of the POVMs implies that $\sum_k a(k|i) = \tr (A_i) $ and $\sum_l b(l|i) = \tr (B_i)$. From this we see that if operators $A_i$ or $B_i$ are traceless, then the quantities $a(k|i)$ or $b(l|i)$ will be all zero, and will not contribute to the sum in Eq. (\ref{lhv}). Hence we may ignore those terms and rewrite Eq. (\ref{lhv}) as:
\begin{eqnarray}
\tr(\rho_{AB} M_k \otimes N_l) = \sum_{i:\tr\{A_i\},\tr\{B_i\}>0} q_i \tr (A_i) \tr (B_i)  {a(k|i) \over \tr(A_i)  } {b(l|i) \over \tr(B_i) } = \sum_{i:\tr\{A_i\},\tr\{B_i\}>0} p_i r(k|i) s(l|i) \label{lhv2}
\end{eqnarray}
where $p_i := q_i \tr (A_i) \tr(B_i), r(k|i):=a(k|i)/\tr (A_i), s(l|i):=b(l|i)/\tr(B_i)$. From the fact that $\tr(\rho_{AB})=1$ it follows that $p_i$ is a normalised probability distribution, and $r(k|i)$ and $s(l|i)$ are normalised conditional probability distributions. Hence we see that Eq. (\ref{lhv2}) supplies a LHV model for the probability distribution $\tr(\rho_{AB} M_k \otimes N_l)$. Note that as demonstrated by the Werner states \cite{W89} in the case of projective quantum measurements, a lack of generalised separability for a given class of measurements does not necessarily imply that a state is non-local w.r.t. those measurements (although for a small enough number of measurements and measurement outcomes the two are equivalent \cite{Shash1}).

In order for a separable decomposition to be useful for constructing LHV models in this way, we would like the local operators appearing in the decomposition to be generalised positive for as large a set of measurements as possible. In this respect there are some separable decompositions that are more useful than others. We will illustrate this point, as well as the theorems that we present later, in reference to three example separable decompositions (all of which are actually examples of \ref{example}) of the density matrix corresponding to the canonical Bell state $\ket{\phi^+}={1 \over \sqrt{2}}(\ket{00}+\ket{11})$:
\begin{enumerate}
\item It is well known from the stabiliser formalism that the Bell state can be written a in terms of the separable decomposition:
\begin{eqnarray}
\ket{\phi^+}\bra{\phi^+}= \frac 14 I \otimes I + \frac 14 \sigma_x \otimes \sigma_x  + \frac 14 \sigma_y \otimes \sigma^T_y +  \frac 14 \sigma_z \otimes \sigma_z,
\end{eqnarray}
where $ \{\sigma_x,\sigma_y,\sigma_z\} $ are the conventional Pauli operators. While the operators $\sigma_x,\sigma_y,\sigma_z$ can be generalised positive for a large class of measurement operators (e.g. measurement operators with Bloch vectors in the positive orthant), this decomposition is useless for constructing LHV models, as the traceless operators $\sigma_x,\sigma_y,\sigma_z$ cannot be generalised positive for a {\it complete} POVM other than the trivial identity measurement.
\item Consider now a second separable decomposition: 
\begin{eqnarray}
\ket{\phi^+}\bra{\phi^+}= \frac 14 (\sqrt{2}\ket{0}\bra{0}) \otimes (\sqrt{2}\ket{0}\bra{0}) +\frac 14 (\sqrt{2}\ket{1}\bra{1}) \otimes (\sqrt{2}\ket{1}\bra{1})+\frac 14\sigma_x \otimes \sigma_x + \frac 14\sigma_y \otimes \sigma^T_y.
\end{eqnarray}
This decomposition better than the first, but still only yields a LHV model only for trivial local $\sigma_z$ observable measurements. By following the procedure given above, the LHV terms for the $ \sigma_z $ measurement are
\begin{align*}
p_i &= \left(\frac 12, \frac 12, 0, 0\right),\\
r(k=+1|i)&=s(k=+1|i)=(1,0,0,0),\\
r(k=-1|i)&=s(k=-1|i)=(0,1,0,0).
\end{align*}
\item Consider now another decomposition of the same Bell state that occurs in the study of discrete Wigner functions (\cite{PhasePointOps,AJRVsmall}):
\begin{align}
\ket{\phi^+}\bra{\phi^+} = \frac 14 \sum_{i=1}^4 W_i \otimes W_i^T, \label{DWFdecomp}
\end{align}
where the $\{W_i\}$ form a subset of the \textit{phase-point operators} \cite{PhasePointOps}, defined as
\begin{align*}
W_1 &= \frac 12(\mathbb{1} + \sigma_x + \sigma_y+\sigma_z),\\
W_2 &= \frac 12(\mathbb{1} + \sigma_x - \sigma_y-\sigma_z),\\
W_3 &= \frac 12(\mathbb{1} - \sigma_x + \sigma_y-\sigma_z),\\
W_4 &= \frac 12(\mathbb{1} - \sigma_z - \sigma_y+\sigma_z).
\end{align*}
The decomposition in Eq. (\ref{DWFdecomp}) is much better than the first two - it yields a LHV model for a much larger set $ \mathcal{S} $ of local measurement operators. For system A, this set turns out to be the measurement operators with Bloch vectors in $\mathcal Q \cap \mathcal T$, where $\mathcal Q$ is the set of single qubit operators (i.e. the Bloch sphere) and $\mathcal T$ is the tetrahedron whose four vertices are the $W_i$ (see Ref. \cite{AJRVsmall}). For system B, the set of local measurement operators is the transpose of those for A.
The set $\mathcal S$ includes all the Pauli observables. In addition, it includes local non-stabilizer POVMs such as $\{c\ket m \bra m, \mathbb{1} - c \ket m \bra m\}$ for small enough $c>0$  where $\ket{m}$ is the (magic \cite{BK}) state with Bloch vector $(1,1,1)/\sqrt 3$.
\end{enumerate}
These examples demonstrate the value of finding a separable decomposition with local operators that are generalised positive for the largest possible classes of local measurements or processes. How may we identify decompositions with such properties?
In order to address this question we will follow the strategy described in our previous work \cite{AJRVsmall}, which was built upon a key property implicitly shared by all notions of generalised positivity: because operators satisfying a given notion of generalised positivity form a convex set, if a set of local operators ${\mathcal{V}}_A$ is positive for a given notion of generalised positivity, then so is the convex hull of any subset ${\mathcal{W}}_A \subset {\mathcal{V}}_A$. This implies that if two separable decompositions exist such that the local convex \textit{state spaces} of one decomposition are contained within those of the other decomposition, then the smaller decomposition can only be more useful. By minimising the local state spaces, for example, we can only enlarge the corresponding dual space of local measurements, thereby achieving better LHV models. This is the motivation behind the question that was posed in \cite{AJRVsmall} and that we continue to explore in this work: what are the {\it smallest} sets of local operators such that a given entangled quantum state $\rho_{AB}$ is separable with respect to them?
\bs

\textit{Problem variants---}  To make this question precise we must define the term `smallest' more precisely. In a previous article \cite{AJRVsmall} we considered three variants of the problem motivated by different definitions of the word `smallest', and presented solutions for some states. In this article we generalise those arguments, obtaining solutions for all bipartite states for two of the problems, and for certain families of bipartite state for the other one. As a technical tool to facilitate the discussion we will also add another variant (which we call Problem 2*)---while this variant was in fact considered in \cite{AJRVsmall}, it was not explicitly defined. Hence there are four problems that we consider, which are:
\begin{itemize}
\item {\bf Problem 1:} For a given bipartite quantum state $\rho_{AB}$ consider all pairs of convex sets $({\cal{V}}_A, {\cal{V}}_B)$ (which we call `state spaces') for which $\rho_{AB}$ is separable, and work out the minimum (in all cases considered in this paper the minimum exists) value of $\|{\cal{V}}_A\|\|{\cal{V}}_B\|$, where we define the size $\|{\cal{V}}\|$ of a state space ${\cal{V}}$ relative to some choice of norm $\|\bullet\|$ as $\|{\cal{V}}\| := \sup \{\|X\||X \in {\cal{V}} \}$. In \cite{AJRVsmall} this was shown to be equivalent to computing a form of cross-norm \cite{OR1,OR2} built from $\|\bullet\|$ (readers that are not familiar with cross norms may see Eq. (\ref{lamdagamma}) for a definition of the versions that we consider here). Depending upon the choice of norm, this problem may not have strong physical meaning by itself. However, we use it as a mathematical tool to solve other variants of the problem.
\item {\bf Problem 2:} For a given quantum state $\rho_{AB}$, consider all pairs of state spaces $({\cal{V}}_A, {\cal{V}}_B)$ for which $\rho_{AB}$ is separable, and identify pairs $({\cal{V}}_A, {\cal{V}}_B)$ such that no strict subsets $({\cal{W}}_A \subset {\cal{V}}_A, {\cal{W}}_B \subseteq {\cal{V}}_B)$ or $({\cal{W}}_A \subseteq {\cal{V}}_A, {\cal{W}}_B \subset {\cal{V}}_B)$ exist for which $\rho_{AB}$ is $({\cal{W}}_A, {\cal{W}}_B)$-separable. The motivation for this problem is the fact that generalised positivity of a set implies generalised positivity of any subset, so it is in our advantage to look for the smallest separable decompositions. As shown in \cite{AJRVsmall}, all the separable decompositions given above for $\ket{\phi^+}\bra{\phi^+}$ are solutions to Problem 2 (the fact that two of them are not of use for constructing LHV models is a motivation for the next two problems).
\item {\bf Problem 2*:} The same as Problem 2, but with the additional constraint that we identify smallest state spaces under the restriction that they must contain the {\it local quantum states} and they only contain operators with unit trace. For the purposes of this article the motivation for Problem 2* is primarily technical, because it turns out that the conic hull \footnote{A conic hull of a set $H$ is the set of all $ah_1+bh_2$ where $a,b>0$ and $h_1,h_2 \in H$, see \cite{Boyd} for details.} of solutions to Problem 2* automatically give a solution our next problem (see \cite{AJRVsmall}). However, in applications of the separable decompositions to classically efficient simulations of quantum systems (not just LHV models), the restriction to unit trace operators can be important for making sure that sampling occurs correctly \cite{AJRVpeps}.
\item {\bf Problem 3:} The same as Problem 2, but with the additional constraint that the state spaces $({\cal{V}}_A, {\cal{V}}_B)$ must be convex {\it cones} of positive trace operators that contain the quantum states (in this case the weights appearing in the separable decomposition need only be positive---whether they form a normalised probability distribution is irrelevant as we are now dealing with cones). The motivation for this problem is explained in \cite{AJRVsmall} and footnote \cite{conelhv}.
\end{itemize}

In the next few sections we present a solutions to these problems for more general states than were considered in \cite{AJRVsmall}. The key technique is the use of certain norms to characterise the local operators and the entangled state. Loosely speaking, the strategy of our arguments follows the reasoning used in  \cite{AJRVsmall}: by evaluating a type of cross-norm (defined later) on an entangled state we obtain an achievable lower bound on the size of the local operator spaces appearing in the decomposition. We are able to generalise \cite{AJRVsmall} to much larger families of states by considering a broader family of norms than were considered there.

\bs

\section{Linearly transformed 2-norms and cross-norms} \label{ournorms}

In this article we will use a particular class of norms on a single state space. These norms are given by a 2-norm of the output of a fixed invertible linear transformation $\Lambda$ on the operator from the state space, and we denote them in the following way:
\begin{equation}
\|X\|_{\Lambda}:= \|\Lambda(X)\|_2.
\end{equation}
Here $\Lambda$ is an invertible linear transformation, and $\|\Lambda(X)\|_2$ refers to the 2-norm
of $\Lambda(X)$ (the 2-norm of an operator $Y$ is defined as $\sqrt{\tr(Y^{\dag}Y)}$). As described above this norm can induce a norm on sets ${\cal{V}}$ of operators via the definition
\begin{equation}
\|{\cal{V}}\|_{\Lambda} := \sup \{\|X\|_{\Lambda} |X \in {\cal{V}}\}.\nonumber
\end{equation}
There are two reasons why these norms are useful to us: firstly the triangle inequality is {\it strict} for them, i.e.
\begin{equation}
\|X + Y\|_{\Lambda} < \|X\|_{\Lambda} + \|Y\|_{\Lambda}, \label{stricttriangle}
\end{equation}
unless $X=cY$ for some positive number $c$ \cite{strictproof}, and secondly we can calculate the resulting cross norms (see e.g. \cite{OR1,OR2,JK} for use in entanglement theory, or Eq. (\ref{lamdagamma}) for the ones we use here) explicitly in a number of cases. We note that strictness of the triangle inequality holds for a variety of other norms (such as the so-called Schatten $p$-norms with $1 < p < \infty$), so it is possible that our arguments could generalise as long as the resulting cross norms can be calculated.

We will use these single state space norms to construct types of cross norm on more than one state space. Given two invertible linear transformations $\Lambda,\Gamma$ we will define the $\Lambda,\Gamma$ cross-norm as:
\begin{equation}
\|\rho_{AB}\|_{\Lambda,\Gamma} := \inf \{\sum_i \|A_i\|_{\Lambda} \|B_i\|_{\Gamma} \, : \,  \rho_{AB} = \sum_i A_i \otimes B_i  \} \label{lamdagamma}
\end{equation}
The arguments presented in \cite{AJRVsmall} show that $\|\rho_{AB}\|_{\Lambda,\Gamma}$ is equivalent to the infimum value of $\|{\cal{V_A}}\|_{\Lambda} \|{\cal{V_B}}\|_{\Gamma}$ over all pairs of sets ${\cal{V_A}},{\cal{V_B}}$ such that $\rho_{AB}$ is $(\cal{V_A},\cal{V_B})$-separable.
We will use these cross norms to help us apply solutions presented in \cite{AJRVsmall} for the maximally entangled state to more general settings. It will be important to our arguments that we can explicitly calculate optimal decompositions (\ref{lamdagamma}) in a variety of cases, thereby adding to solutions that have been computed in previous works \cite{OR1,OR2}.

\section{Summary of Results} \label{results}

In this section we summarize the main results of the paper. We describe the linear transformations $\Lambda, \Gamma$ by expanding in the orthonormal basis that describes the operator-Schmidt decomposition of the state $\rho_{AB}$ of interest. So for instance we write:
\begin{equation}
\Lambda (X_i) = \sum_{i,j} R_{ij} X_j,
\end{equation}
where $R=\{R_{ij}\}$ is a $D \times D$ matrix. With this notation our main results are as follows.

\begin{theorem} \label{th1} (Solutions to Problem 1)
Consider a bipartite quantum state $\rho_{AB}$ with operator-Schmidt decomposition $\rho_{AB}=\sum^D_{i=1} s_i X_i \otimes Y_i$ with $s_i > 0$. Define $S = \mathrm{diag}(s_i)$ and make arbitrary choices of the following: an invertible $D \times D$ positive diagonal matrix $R$, positive constants $c_k, k=1,\dots,N$, probability distribution $p_k, k=1,\dots,N$, and $D \times N$ isometry $U$. Then
\begin{align}
\rho_{AB} = \sum^N_{k=1} p_k A^k \otimes B^k,
\end{align}
with
\begin{eqnarray}
A^k &=&  \sqrt{{1 \over p_k c_k}} \sum^D_{i=1}  (\sqrt{S} R^{-1} U )_{ik} X_i, \nonumber \\
B^k  &=& \sqrt{{ c_k \over p_k }} \sum^D_{i=1} (\sqrt{S} R U )^*_{ik} Y_i,
\end{eqnarray}
gives a separable decomposition for $\rho_{AB}$ that solves Problem 1, i.e. it achieves the minimal $R,R^{-1}$ cross-norm. Moreover, all $R,R^{-1}$ cross-norm achieving separable decompositions can be written in this form. By choosing $R=\sqrt{S}$, $U_{ij} = \delta_{ij}$, $c_k = p_k$, we find that the operator-Schmidt decomposition itself is a solution to Problem 1.
\end{theorem}

In the examples given for the Bell state in the introduction all the decompositions satisfy this theorem and are hence solutions to Problem 1. More generally, the operator-Schmidt decomposition in Eq. \eqref{example} satisfies this theorem.

\begin{theorem} \label{th2} (Solutions to Problem 2)
Consider a bipartite quantum state $\rho_{AB}$ with operator-Schmidt decomposition $\rho_{AB}=\sum^D_{i=1} s_i X_i \otimes Y_i$ with $s_i > 0$. Define $S = \mathrm{diag}(s_i)$ and make arbitrary choices of the following: an invertible $D \times D$ positive diagonal matrix $R$, a positive constants $c$ and $D \times D$ unitary matrix $U$. Then
\begin{align}
\rho_{AB} = \sum^N_{k=1} p_k A^k \otimes B^k,
\end{align}
with
\begin{eqnarray}
A^k =  \sqrt{{1 \over p_k c}} \sum^D_{i=1}  (\sqrt{S} R^{-1} U )_{ik} X_i, \nonumber \\
B^k =  \sqrt{{ c\over p_k }} \sum^D_{i=1}  (\sqrt{S} R U )^*_{ik} Y_i, \nonumber \\
p_k =  \sum^D_{i=1} |U_{ik}|^2 {s_i \over \sum_j s_j},
\end{eqnarray}
gives a separable decomposition for $\rho_{AB}$ that solves Problem 2, i.e. it achieves the minimal $R,R^{-1}$ cross-norm and is a smallest one, in the sense that the sets:
\begin{eqnarray}
{\mathcal{V}}_A := {\rm{conv}}\{A^k |k=1,\dots,D\}, \nonumber \\
{\mathcal{V}}_B := {\rm{conv}}\{B^k |k=1,\dots,D\}, \nonumber
\end{eqnarray}
cannot be made smaller while continuing to admit a separable decomposition for $\rho_{AB}$. By choosing $R=\sqrt{\frac 1c}\mathbb{1} $ and $U_{ij} = \delta_{ij}$ we find that the operator-Schmidt decomposition itself is a solution to Problem 2.
\end{theorem}
Note the close parallels between this result and theorems on ensemble decompositions of density matrices (see \cite{KL04,HKLW07,HJW93} and Theorem 2.6 in \cite{NC01}).

In the examples given for the Bell state in the introduction all decompositions are satisfy this theorem and are hence solutions to Problem 2. More generally, the operator-Schmidt decomposition in Eq. \eqref{example} satisfies this theorem and thus provides a solution to Problem 2.

\begin{theorem} \label{th3} (Solutions to Problem 2* and Problem 3 for some states.)
Consider a bipartite quantum state that can be expressed as $\rho_{AB} = {\cal{E}}\otimes {\cal{F}}(\Psi)$, where $\Psi$ is the canonical maximally entangled state and ${\cal{E}},{\cal{F}}$ are invertible linear maps. Then provided ${\cal{E}},{\cal{F}}$ satisfy the following:
\begin{itemize}
\item[(A)] a basis of operators $W_i$ can be found such that $\tr(W^{\dag}_i W_j) = d \delta_{ij}$ and $\tr({\cal{E}}({W}_i))=1$ and $\tr({\cal{F}}(W^T_i))=1$,
\item[(B)] $||{\cal{E}}^{-1}(\sigma)||_2, ||{\cal{F}}^{-1}(\sigma)||_2 < \sqrt{d}$ for any local quantum states $\sigma$,
\end{itemize}
then $\rho_{AB}$ is separable w.r.t both the unit trace convex hulls
\begin{equation}
\mathrm{conv}\{{\cal{E}}(W_i) \bigcup {\mathcal{Q}}_A\} \,\,\, , \,\,\,
\mathrm{conv}\{{\cal{F}}(W^T_i) \bigcup {\mathcal{Q}}_B\}
\end{equation}
where ${\mathcal{Q}}_A,{\mathcal{Q}}_B$ are the local quantum states for particles $A$, $B$ respectively, and the positive trace conic hulls:
\begin{equation}
\mathrm{coni}\{{\cal{E}}(W_i) \bigcup {\mathcal{Q}}_A\} \,\,\, , \,\,\,
\mathrm{coni}\{{\cal{F}}(W^T_i) \bigcup {\mathcal{Q}}_B\}
\end{equation}
Moreover, these convex (resp. conic) sets contain no convex (resp. conic) strict subsets that both contain the local quantum states and maintain separability of $\rho_{AB}$. Hence they give a solution to Problem 2* (resp. Problem 3) for any $\rho_{AB}$ satisfying the two conditions. Condition (A) holds for all bipartite pure states, and all bipartite states that can be obtained from $\Psi$ by local CPTP maps. Condition (B) holds for all bipartite states satisfying $\min_k s_k > 1/d^2$, where $s_k$ are the operator-Schmidt coefficients, which in turn holds for states in a finite region around the maximally entangled state.
\end{theorem}

In the examples given for the Bell state in the introduction, taking the convex/conic hull of the local quantum states with the local operators of the last decomposition gives an example of this solution.

\section{Proof of Theorem 1: Characterisation of solutions to Problem 1 for some cross norms.} \label{characterisation}

In this section we will characterise solutions to Problem 1 for a particular subset of the cross norms defined by picking $\Gamma=\Lambda^{-1}$. We choose these norms because they enable us to treat the problem in a very similar way to the problem of finding decompositions of positive operators, and hence draw upon solutions to that problem (see \cite{KL04,HKLW07,HJW93}, or the discussion around Theorem 2.6 in \cite{NC01}). In later sections we will show that among the solutions we obtain to  Problem 1 there are many that also solve Problem 2.

We begin by deriving a simple lower bound on the cross norms defined by picking $\Gamma=\Lambda^{-1}$. Consider an operator written out in an operator-Schmidt decomposition:
\begin{eqnarray}
\rho_{AB} =  \sum^D_{i=1} s_i X_i \otimes Y_i \equiv \sum_i s_i \delta_{ij} X_i \otimes Y_j.
\end{eqnarray}
Now consider any generalised separable decomposition of $\rho_{AB}$:
\begin{eqnarray}
\rho_{AB} =  \sum^N_{k=1} p_k A^k \otimes B^k.
\end{eqnarray}
We will expand $A^k$ and $B^k$ in terms of the operators $X_i, Y_j$ in the following way: $A^k = \sum a^k_i X_i$ and $B^k = \sum (b^k_i)^* Y_i$, where $a^k_{i}$ and $b^k_{i}$ are expansion coefficients (the complex conjugate on the $b^k_i$ is for later convenience). We will perform all calculations in this representation. Let us use the symbols $\boldsymbol{a}^k$ and $\boldsymbol{b}^k$ to represent vectors of these expansion coefficients $a^k_{i}$ and $b^k_{i}$, respectively. It is useful to note that the the 2-norms of the local operators are given by $\|A^k\|_2 = \|\boldsymbol{a}^k\|_2$ and $\|B^k\|_2 = \|\boldsymbol{b}^k\|_2$. For the two expressions for $\rho_{AB}$ to be equal, we require the following three equivalent equations to hold:
\begin{eqnarray}
&& s_i \delta_{ij} = \sum^N_{k=1} p_k \alpha^k_i (\beta^k_j)^*, \nonumber \\
\Leftrightarrow && \sqrt{s_i} \sqrt{s_j} \delta_{ij}= \sum^N_{k=1} p_k \alpha^k_i (\beta^k_j)^*,  \nonumber \\
\Leftrightarrow && \delta_{ij}= \sum^N_{k=1} p_k {\alpha^k_i \over \sqrt{s_i}} \left({\beta^k_j \over \sqrt{s_i}} \right)^*.  \label{general}
\end{eqnarray}
If we sum the first of these equations over $i=j$ we get:
\begin{eqnarray}
\sum_i s_i = \sum_k p_k \langle \boldsymbol{b}^k, \boldsymbol{a}^k\rangle \leq \sum_k p_k\|\boldsymbol{a}^k\|\|\boldsymbol{b}^k\|, \label{normbound1}
\end{eqnarray}
where the inequality follows from Cauchy-Schwarz inequality. This equation gives us a lower bound for the
usual cross 2-norm, which is in fact tight \cite{OR1,OR2}. However, the route taken to this equation can be used to give a lower bound for the larger class of norms involving linear transformations $\Lambda=\Gamma^{-1}$ on the two subsystems. To see this consider representing $\Lambda$ by a linear transformation $R$ acting on the space of vectors $\boldsymbol{a}^k$. Then we note that:
\begin{eqnarray}
\langle \boldsymbol{b}^k, \boldsymbol{a}^k\rangle = \langle R^{-1}\boldsymbol{b}^k, R \boldsymbol{a}^k\rangle. \nonumber
\end{eqnarray}
Given $R$ we may rewrite Eq. (\ref{normbound1}) as:
\begin{eqnarray}
\sum_i s_i &=& \sum_k p_k \langle R^{-1} \boldsymbol{b}^k, R \boldsymbol{a}^k\rangle, \nonumber\\
 &\leq& \sum_k p_k \|R \boldsymbol{a}^k\| \|R^{-1}\boldsymbol{b}^k\|. \label{normbound2}
\end{eqnarray}
This is the lower bound that we set out to derive. We may consider $\|R \boldsymbol{a}^k\|, \|R^{-1}\boldsymbol{b}^k\|$ as defining a cross norm calculated by first applying invertible linear transformations $R$ to the first system, and $R^{-1}$ to the second system, and then computing the 2-norm. In other words we are considering a cross norm of the type discussed above, but with $\Lambda$ set by $R$ and $\Gamma$ set by $R^{-1}$.

Although the lower bound (\ref{normbound2}) will not be tight in general, we will now show that it is true if we restrict $R$ that to be diagonal and positive. In other words we will demonstrate that for diagonal positive $R$ the reverse inequality holds too:
\begin{eqnarray}
\sum_k p_k \|R \boldsymbol{a}^k\| \| R^{-1} \boldsymbol{b}^k\|  \leq \sum_i s_i \label{assertion1}.
\end{eqnarray}
The Cauchy-Schwarz inequality implies that the only way that the previous two equations could possibly be compatible is if we have $\|R \boldsymbol{a}^k\|_2 \| R^{-1} \boldsymbol{b}^k\|_2=\langle R^{-1} \boldsymbol{b}^k, R \boldsymbol{a}^k\rangle$ for all $k$, and hence $R \boldsymbol{a}^k$ and $R^{-1} \boldsymbol{b}^k$ must be proportional:
\begin{eqnarray}
&& R^{-1} \boldsymbol{b}^k =  c_k R \boldsymbol{a}^k,  \nonumber \\
\Rightarrow &&  \boldsymbol{b}^k = c_k R^2 \boldsymbol{a}^k, \label{linktopos}
\end{eqnarray}
for some positive numbers $c_k$ that may depend upon $k$. As we now show, it is this condition that links our problem to previously known results on the decomposition of positive operators. Let us insert Eq. (\ref{linktopos}) into one of Eqs. (\ref{general}) to eliminate the $b^k_j$:
\begin{eqnarray}
s_i \delta_{ij} = \sum_k p_k c_k a^k_i \sum_g \left( R^2 \right)_{jg} (a^k_g)^*.
\end{eqnarray}
Multiplying both sides by $R^{-2}$ gives:
\begin{eqnarray}
&& \sum_j s_i \delta_{ij} \left( R^{-2} \right)_{hj}  = \sum_k p_k c_k a^k_i (a^k_h)^*, \nonumber \\
\Rightarrow && \sum_j s_i \delta_{ij} \left( R^{-2} \right)_{jh} = \sum_k p_k c_k a^k_i (a^k_h)^*. \label{indexform}
\end{eqnarray}
We are looking for $p_k,c_k$ (which are positive) and $a^k_i$ that satisfy these equations. Hence we see from Eq. (\ref{indexform}) that we are looking for a positive decomposition of the matrix on the left-hand-side, and may thereby draw upon previous results on that problem (\cite{NC01,HJW93,KL04,HKLW07}). It would of course not be possible to do this if the operator on the left-hand-side were not positive, and this is why we picked $R$ deliberately to be diagonal and positive so that the LHS is positive (although there is a slightly more general class of $R$ that would satisfy this, it turns out that the more general argument still leads to the same decompositions). If we define a $D \times D$ matrix $S$ as the diagonal matrix with matrix elements $s_i \delta_{ij}$, and another $D \times N$ matrix $Z$ with matrix elements $Z_{ik} = \sqrt{p_k c_k} a^k_i$, then Eq. (\ref{indexform}) can be rewritten more concisely as:
\begin{eqnarray}
S R^{-2} = Z Z^{\dag}. \label{matrixform}
\end{eqnarray}
The separable decomposition is represented by the matrix $Z$ and the coefficients $p_k, c_k$. We would hence like to find values for these variables that solve this equation. Notice that Eq. (\ref{matrixform}) is simply the statement that $Z$ is a square root of the positive operator $S R^{-2}$. Hence:
\begin{eqnarray}
Z = \sqrt{S} R^{-1} U, \label{zed}
\end{eqnarray}
for an isometry $U$.

Let us summarise where we are. Given a positive invertible diagonal matrix $R$, a probability distribution $p_k$, and a set of positive numbers $c_k$, a separable decomposition satisfying the assertion (\ref{assertion1}) can exist only if $Z$ is given by (\ref{zed}). Writing these equations out explicitly, we find that any separable decomposition achieving (\ref{assertion1}) must {\it necessarily} be of the form:
\begin{eqnarray}
&& \rho_{AB} = \sum^N_{k=1} p_k A^k \otimes B^k, \nonumber \\
A^k = \sum a^k_i X_i &=&  \sqrt{{1 \over p_k c_k}} \sum_i  (\sqrt{S} R^{-1} U )_{ik} X_i, \nonumber \\
B^k = \sum (b^k_i)^* Y_i &=& \sqrt{{ c_k \over p_k }} \sum_i (\sqrt{S} R U )^*_{ik} Y_i. \label{separable}
\end{eqnarray}
To show that this is also sufficient, we must verify that this is indeed a separable decomposition (the fact that it satisfies (\ref{assertion1}) is by construction). A short calculation verifies that this is true for arbitrary $p_k,c_k >0$ \cite{sepverify}.
Hence, we see that Eq. (\ref{separable}) is indeed a separable decomposition for $\rho_{AB}$, and moreover it characterises all separable decompositions that achieve the cross-norm defined by $R,R^{-1}$. The $c_k$ simply serves as a `scale' factor that cancels out when forming $A^k \otimes B^k$---so we can use it to make the norm of any operator on Bob's side small, but only at the expense of making the corresponding operator on Alice's side large. Note that (as follows from the connections in  \cite{AJRVsmall}) Eqs. (\ref{separable}) characterise all cross-norm achieving decompositions, not just separable ones (indeed nothing in our above arguments requires $\sum_k p_k=1$).

\section{Proof of Theorem 2: decompositions with a minimal number of operators having the same norm} \label{Prob2}

The previous section characterises all separable decompositions achieving the minimal value of the cross-norm defined through $R$ and $R^{-1}$. We will now show that some of the decomposition of Theorem \ref{th1} also yield solutions to Problem 2. We will do this by picking decompositions with two properties: (a) for which there is the minimal number of operators ($N=D$) on each side, (b) for which the local expansion vectors on Alice's side all have the same norm, and for which the local expansion vectors on Bob's side all have the same norm. In such cases, the only operators from within the local state spaces that have a high enough norm to be part of a separable decomposition are hence the extremal points (the fact that the triangle inequality is strict means that {\it only} the the extremal points have a high enough norm). Moreover, as we need all $D$ of them from each local state space, such state spaces have the property that we cannot make them smaller while preserving separability.

To complete the argument we hence need to show how to pick decompositions with the two properties (a),(b). Property (a) can be guaranteed by restricting the $U$ appearing in the decompositions to be a square unitary matrix. This means that we only have decompositions with the minimal number $D$ of local operators, hence satisfying (a). We must now work out which of these decompositions satisfy (b), i.e. with all operators having the same norm. The squares of the norms of the vectors $\boldsymbol{a}^k, \boldsymbol{b}^k$ in the decomposition are given by:
\begin{eqnarray}
\| R \boldsymbol{a}^k\|^2 =  {1 \over p_k c_k} \sum_i |U_{ik}|^2s_i, \nonumber \\
\| R^{-1} \boldsymbol{b}^k\|^2 =  {c_k \over p_k } \sum_i |U_{ik}|^2 s_i.
\end{eqnarray}
If we wish $\| R \boldsymbol{a}^k\|^2$, say, to be same for all $k$ then we require it to be a constant, which we denote $w$:
\begin{eqnarray}
w =   {1 \over p_k c_k} \sum_i |U_{ik}|^2 s_i,
\end{eqnarray}
i.e. we require that:
\begin{eqnarray}
 w p_k c_k =  \sum_i |U_{ik}|^2 s_i. \label{constant}
\end{eqnarray}
Note that the $|U_{ik}|^2$ form the entries of a doubly stochastic matrix (note that it is square as $N=D$, the minimal number allowed). The value of $w$ is fixed by summing both sides over $k,i$:
\begin{eqnarray}
 w  =  {\sum_i s_i \over \sum_k p_k c_k}.
\end{eqnarray}
Hence Eq.~(\ref{constant}) becomes:
\begin{eqnarray}
{ p_k c_k \over \sum_l p_l c_l} =  \sum_i |U_{ik}|^2 {s_i \over \sum_j s_j}.
\end{eqnarray}
We may go through the same argument for the operators on Bob's side. The result is obtained by replacing $c_k$ with $1/c_k$:
\begin{eqnarray}
{ p_k/c_k \over \sum_l p_l/c_l} =  \sum_i |U_{ik}|^2 {s_i \over \sum_j s_j}.
\end{eqnarray}
These two equations tell us that the probability distributions defined by:
\begin{eqnarray}
{ p_k c_k \over \sum_l p_l c_l} \,\,\,\,;\,\,\,\,{ p_k/c_k \over \sum_l p_l/c_l},\label{twoprobs}
\end{eqnarray}
must be majorized by the probability distribution defined by:
\begin{eqnarray}
{s_i \over \sum_j s_j},
\end{eqnarray}
with the corresponding doubly stochastic matrix (taking the latter distribution to the former distributions) given by the absolute square elements of a unitary:
\begin{eqnarray}
 |U_{ik}|^2.
\end{eqnarray}
The only way that this can happen is if the two probability distributions (\ref{twoprobs}) are identical, which requires that for all $k,k'$:
\begin{eqnarray}
{ p_k c_k \over p_k/c_k} = { p_{k'} c_{k'} \over p_{k'}/c_{k'}}, \nonumber \\
\Rightarrow c_k = c_{k'}.
\end{eqnarray}
So we must set $c_k=c$ where $c$ is some positive constant, and hence in Eq. (\ref{twoprobs}) the $c$'s cancel and the two probability distributions become identical. Hence, we find that a separable decomposition:
\begin{eqnarray}
\rho_{AB} = \sum^D_{k=1} p_k A^k \otimes B^k, \nonumber \\
A^k =  \sqrt{{1 \over p_k c}} \sum^D_{i=1}  (\sqrt{S} R^{-1} U )_{ik} X_i, \nonumber \\
B^k =  \sqrt{{ c\over p_k }} \sum^D_{i=1}  (\sqrt{S} R U )^*_{ik} Y_i, \label{prob2soln}
\end{eqnarray}
is a smallest one if the $p_k$ are chosen so that:
\begin{eqnarray}
 p_k = \sum^D_{i=1}  |U_{ik}|^2 {s_i \over \sum_j s_j}.
\end{eqnarray}
The above argument can also be used to construct solutions to Problem 2 under the restriction that the local state spaces are Hermitian. As a quantum state $\rho_{AB}$ is Hermitian, the $X_i$ and $Y_i$ in the operator-Schmidt decomposition can be taken to be Hermitian. If we want to restrict the separable decomposition to involve only Hermitian operators, then we can do so by further restricting the unitary $U$ to in fact be a real orthogonal matrix $O$.

Note the close parallels between this result and theorems on ensemble decompositions of density matrices (see \cite{KL04,HKLW07,HJW93} and Theorem 2.6 in \cite{NC01}). By choosing $R=\sqrt{\frac 1c}\mathbb{1} $ and $U_{ij} = \delta_{ij}$ we find that the operator-Schmidt decomposition itself is a solution to Problem 2, as we claimed in the introduction.

\section{Proof of Theorem 3: Obtaining solutions to Problem 3 from the maximally entangled case} \label{prob3}

The solutions to Problems 1 and 2 in the previous sections do not provide solutions to Problem 2* or Problem 3. The motivation for Problem 3 is that its solution maximises the set of quantum effects for which the separable decomposition gives a LHV model: taking the conic hull of a state space does not reduce the dual set of quantum measurements for which the state space is positive; requiring the inclusion of the quantum states in the the conic hull means that the dual only contains quantum \textit{effects} (that is, positive semidefinite operators); and requiring positive trace means that all such quantum effects can be completed to give POVMs (in the sense that for positive trace operators, positivity w.r.t. quantum POVM element $M$ automatically implies positivity w.r.t. $I-M$ too).

In order to solve Problem 3, we must identify conic state spaces containing the local quantum states that cannot be made smaller while maintaining separability of $\rho_{AB}$. To do this we will first identify, for a family of bipartite quantum states that includes all pure states, local state spaces consisting of unit trace operators that solve Problem 2. We will then argue that, provided that the quantum state $\rho_{AB}$ is correlated enough, adding the unit trace local quantum operators to these state spaces also solves Problem 2*. The conic hull of these state solutions automatically gives solutions to Problem 3, as explained in \cite{AJRVsmall}.

So our main aim is to hence obtain solutions to Problem 2*. One might attempt a solution to Problem 2* by taking the solutions to Problem 2 described in Theorem \ref{th2}, and looking for $U,p_k,c$ that make the operators in Eq. (\ref{prob2soln}) have unit trace. Although \cite{AJRVsmall} shows that was possible to do this for the maximally entangled state, we have not been able to carry out that approach for more general $\rho_{AB}$. However, we will see that it is possible to transform the solutions for the maximally entangled state into solutions to both Problem 2* (and hence Problem 3) for more general quantum states $\rho_{AB}$ that possess two technical properties that we describe below.

Let us first recap the solution to Problem 2 in the case of the maximally entangled state on particles with $d = \dim (\mathcal{H}_A) = \dim (\mathcal{H}_B)$:
\begin{equation}
\ket{\Psi} = {1\over \sqrt{d}} \sum^{d}_i \ket{ii}.
\end{equation}
In  \cite{AJRVsmall} it was shown that {\it any} basis of orthogonal local operators $\{C_i|i=1...d^2\}$ normalised to $\tr(C_i C^{\dag}_j)=d\delta_{ij}$ gives a smallest separable decomposition of $\Psi := \ket{\Psi}\bra{\Psi}$ via:
\begin{equation}
\Psi = {1\over d^2} \sum_i C_i \otimes C^T_i. \label{maxent}
\end{equation}
In order to see how this can be of use for solving Problem 2* for more general $\rho_{AB}$, consider writing $\rho_{AB}$ in terms of local invertible transformations on $\Psi$, i.e.
\begin{equation}
\rho_{AB} = {\cal{E}}\otimes {\cal{F}}(\Psi),
\end{equation}
where ${\cal{E}},{\cal{F}}$ are linear invertible maps. As long as $\rho_{AB}$ has the same operator-Schmidt rank as $\Psi$, i.e. $d^2$, then this is possible. Note that we do not require ${\cal{E}},{\cal{F}}$ to be completely positive or trace preserving maps---they only have to be linear and invertible. Although there are many possible choices for ${\cal{E}},{\cal{F}}$, we will explicitly consider the following ones that are constructed from the Schmidt form of $\rho_{AB}$:
\begin{eqnarray}
{\cal{E(\sigma)}}  = \sum_j \sqrt{s_j} X_j \tr(C^{\dag}_j \sigma),\nonumber \\
{\cal{F(\sigma)}}  = \sum_j \sqrt{s_j} Y_j \tr((C^T_j)^{\dag} \sigma). \label{ourtransforms}
\end{eqnarray}
Let us now consider an alternative separable decomposition for $\Psi$ in terms of another set of basis operators (where $\tr(W^{\dag}_i W_j) = d\delta_{ij}$), i.e. now decompose $\Psi$ as:
\begin{eqnarray}
\Psi = {1 \over d^2} \sum_i W_i \otimes W^T_i.
\end{eqnarray}
For the state $\rho_{AB}$ and for any suitable choice of ${\cal{E}},{\cal{F}}$ we can get a separable decomposition for $\rho_{AB}$ by acting upon this decomposition of $\Psi$ with ${\cal{E}},{\cal{F}}$ to give:
\begin{equation}
\rho_{AB} = {1\over d^2} \sum^{d^2}_{i=1} {\cal{E}}({W}_i) \otimes {\cal{F}}(W^T_i). \label{rhofrompsi}
\end{equation}
It is not difficult to see that this gives a solution to Problem 2: the local state spaces given by the convex hulls of the $\{{\cal{E}}({W}_i)\}$ and the $\{{\cal{F}}(W^T_i)\}$ are solutions to Problem 2 for $\rho_{AB}$ because if they weren't, we could identify smaller subset that give a separable decomposition for $\rho_{AB}$, and then apply the inverse maps to get smaller state spaces than the convex hulls of the $W_i, W^T_i$ for which $\Psi$ is separable. As this is not possible, this means that $\{{\cal{E}}({W}_i)\}$ and $\{{\cal{F}}(W^T_i)\}$ give solutions to Problem 2 for $\rho_{AB}$ of full operator-Schmidt rank.
Note that this method of solution is different to the one developed earlier sections---there is no restriction on ${\cal{E}},{\cal{F}}$ other than invertibility, and so far we have not considered any norms on $\rho_{AB}$ anyway.

We will now see how this method can be extended to solve Problem 2* provided that we make two further assumptions about $\rho_{AB}$:
\begin{itemize}
\item[(A)] There exist $W_i,W_i^T$ satisfying $\tr(W^{\dag}_i W_j) = d\delta_{ij}$ such that the $\{{\cal{E}}({W}_i)\}$ and $\{{\cal{F}}(W^T_i)\}$ are of unit trace. In the appendix we show that an infinite number of such $W$s can be found if $\rho_{AB}$ is such that $\tr(X_i)=\tr(Y_i)$, and this is in turn true for instance if $\rho_{AB}$ is a bipartite pure state. Furthermore, in the case that ${\cal{E}},{\cal{F}}$ are CPTP maps then this can always be achieved by picking $W$s with unit trace.
\item[(B)] There exist ${\cal{E}},{\cal{F}}$ such that ${\cal{E}}^{-1},{\cal{F}}^{-1}$ don't increase 2-norms too much, by which we mean that for all local quantum states $\sigma$ on either Alice or Bob's side, it holds that $||{\cal{E}}^{-1}(\sigma)||_2, ||{\cal{F}}^{-1}(\sigma)||_2 < \sqrt{d}$. In the appendix we show that this is guaranteed provided that $\min_k s_k \geq 1/d^2$, which is in turn true provided for states in a region around the maximally entangled state (because for the maximally entangled state $\min_k s_k = 1/d$).
\end{itemize}
Let us now see why if $\rho_{AB}$ satisfies these two conditions we are able to solve Problem 3. Define a cross norm as follows:
\begin{equation}
\|\rho_{AB}\|_{{\cal{E}}^{-1},{\cal{F}}^{-1}} := \inf \{\sum_i ||{\cal{E}}^{-1}(A_i)||_2 ||{\cal{F}}^{-1}(B_i)||_2 \, : \,  \rho_{AB} = \sum_i A_i \otimes B_i  \}.
\end{equation}
Then from the fact that $\rho_{AB} = {\cal{E}}\otimes {\cal{F}}(\Psi)$ we are able to calculate the value of this norm on $\rho_{AB}$. Indeed we see that:
\begin{equation}
\|\rho_{AB}\|_{{\cal{E}}^{-1},{\cal{F}}^{-1}}= \| \Psi \|_{I,I} = d,
\end{equation}
because if a product decomposition of $\rho_{AB}$ existed giving a lower value of $\|\rho_{AB}\|_{{\cal{E}}^{-1},{\cal{F}}^{-1}}$, then by applying the inverse transformations we would end up with a decomposition for $\Psi$ with a lower value than $d$ for $\| \Psi \|_{I,I}$, which is not possible (note that $\| \Psi \|_{I,I}$ is more conventionally denoted $\| \Psi \|_{\gamma}$ \cite{OR1}).

Now let us consider the unit trace convex hulls of the local quantum states with either $\{{\cal{E}}({W}_i)\}$ or $\{{\cal{F}}(W^T_i)\}$ . We will argue that these state spaces are solutions to Problem 2* by invoking properties (A),(B).
First we note that the operators $\{{\cal{E}}({W}_i)\}$ and $\{{\cal{F}}(W^T_i)\}$ have the property that
$||{\cal{E}}^{-1}({\cal{E}}(W_i))||_2 = ||{\cal{F}}^{-1}({\cal{F}}(W^T_i))||_2= \sqrt{d}$.
Hence, if we consider the unit trace convex hulls made from $\{{\cal{E}}({W}_i)\}$ and $\{{\cal{F}}(W^T_i)\}$ with the local quantum states, then a separable decomposition for $\rho_{AB}$ in terms of these unit trace convex sets can only involve the operators $\{{\cal{E}}({W}_i)\}$ and $\{{\cal{F}}(W^T_i)\}$. This is because by applying the strictness of the triangle inequality (\ref{stricttriangle}), all other operators in these sets have a norm $||{\cal{E}}^{-1}(\bullet)||_2$ or $||{\cal{E}}^{-1}(\bullet)||_2$ that is too small - only the $\{{\cal{E}}({W}_i)\}$ and $\{{\cal{F}}(W^T_i)\}$ can allow us to attain $\|\rho_{AB}\|_{{\cal{E}}^{-1},{\cal{F}}^{-1}}= d$. Moreover, all of the operators $\{{\cal{E}}({W}_i)\}$ and $\{{\cal{F}}(W^T_i)\}$ must appear in the separable decomposition, due to the operator-Schmidt rank of $\rho_{AB}$. This means that the convex hulls of $\{{\cal{E}}({W}_i)\}$ and $\{{\cal{F}}(W^T_i)\}$ with the local quantum states are solutions to Problem 2* for $\rho_{AB}$, and hence the conic hulls of these sets give a solution to Problem 3.

\section{Connections with Decompositions of Linear operators}

There is a standard association between tensor products and linear operators that has appeared in multiple ways in the literature (equivalence of the singular value decompositions and the Schmidt-decomposition, Choi-Jamio\l{}kowski isomorphism, etc). All our results can hence be expressed in terms of linear operators rather than tensor products, and indeed the results on Problems 1 and 2 have a close relationship to previous results on decompositions of positive operators. In this section we clarify these connections.

Following the association made in \cite{OR2}, Eq. (\ref{general}) can be viewed as a decomposition of a linear operator on the left hand side in terms of a ket-bra of vectors $\boldsymbol{a}^k$ and $\boldsymbol{b}^k$ appearing on the right hand side. With this viewpoint, Problem 2 can hence be reexpressed in the following way. Given a linear operator $G$ acting on some finite dimensional complex vector space, find convex decompositions in terms of rank-1 operators:
\begin{equation}
G = \sum_i p_i \ket{A_i}\bra{B_i},
\end{equation}
that are `smallest' ones, in the sense that the convex hulls $\rm{conv}\{\ket{A_i}\}$ and $\rm{conv}\{\ket{B_i}\}$ cannot be made smaller while continuing to admit a rank-1 convex decomposition for $G$. Problem 1 on the other hand asks for convex decompositions into rank-1 operators where the sums of the products of various norms of the $\{\ket{A_i}\}$ and $\{\ket{B_i}\}$ are minimised. The proofs of Theorems 1 and 2 show that some such solutions can be identified by reducing the problem to one of finding decompositions of positive operators, and moreover that the Schmidt/Singular value decomposition and simple transformations of it lead to these solutions. Problem 3 appears to have no similar natural linear algebraic interpretation.

\section{Conclusions}

We have considered the construction of separable decompositions for entangled quantum states that are obtained by relaxing the requirement that the local operators in the decomposition be positive unit trace quantum states. The motivation for this problem is the construction of LHV models and classically efficient simulations for bipartite entangled quantum states, or the multipartite quantum states that are built from them.

In this context it is of interest to find `smallest' decompositions, where the sets of local operators cannot be made smaller while continuing to allow a separable description. We consider four variants of this problem. The first (Problem 1) uses norms to quantify how `small' a set of operators is. The second (Problem 2) uses set inclusion (so that one set is smaller than another if it is contained within it). The third (Problem 2*) and fourth (Problem 3) variants we add restrictions (requiring the sets to be unit trace, contain the quantum states, or be cones) that are important when constructing LHV models.

For the first problem we present all solutions for bipartite states for some norms. For the second problem we present some solutions for all bipartite quantum states. For the remaining two problems we obtain solutions for some bipartite states, including pure states in a region around the maximally entangled state.

Our results generalise those of \cite{AJRVsmall}, and have strong relationships to the study of generalised probabilistic theories with operator spaces \cite{OperatorGPT}, the study of discrete phase space distributions \cite{PhasePointOps}, cross norm entanglement measures \cite{OR1,OR2}, and decompositions of linear operators \cite{HJW93,KL04,NC01,HKLW07}. In the manner of \cite{AJRVpeps} we believe that they may find applications in the study and simulation of entangled many-body quantum states.

\section*{Acknowledgements}
HA acknowledges the financial support of the EPSRC. SJ is supported by an Imperial College London Junior Research Fellowship. SJ also acknowledges ERC grants QFTCMPS, and SIQS by the cluster of excellence EXC 201 Quantum Engineering and Space-Time Research. This work was begun when HA, SJ, and SSV were supported by EPSRC grant EP/K022512/1.

\section{Appendix: Conditions (A) and (B)}

\begin{itemize}
\item[A:] We show that condition (A) can be met if the operators in the operator-Schmidt decomposition of $\rho_{AB}$ satisfy $\tr(X_i)=\tr(Y_i)$. If we take ${\cal{E}},{\cal{F}}$ to be as in Eq. (\ref{ourtransforms}), then we are looking for $W$s that satisfy:
\begin{eqnarray}
1 = \tr({\cal{E}}(W_k)) &=&  \sum_j \sqrt{s_j} \tr(X_j) \tr(C_j^{\dag} W_k), \nonumber \\
&=& \sum_j \sqrt{s_j}d  T_{kj} \tr(X_j), \nonumber \\
1 = \tr({\cal{F}}(W_k)) &=&  \sum_j \sqrt{s_j}\tr(Y_j) \tr((C^T)^{\dag}_j W^T_k), \nonumber \\
&=& \sum_j \sqrt{s_j}d  T_{kj} \tr(Y_j), \nonumber
\end{eqnarray}
where $T_{kj}$ form the elements of a unitary defined by $d T_{kj}=\tr(C^{\dag}_jW_k) = \tr((C^T)^{\dag}_j W^T_k)$. Working directly with $T$ rather than the $W$s, we find that we must solve:
\begin{eqnarray}
{1 \over d} &=&  \sum_j \sqrt{s_j}  T_{kj} \tr(X_j), \nonumber \\
{1 \over d} &=&  \sum_j \sqrt{s_j} T_{kj} \tr(Y_j).  \nonumber
\end{eqnarray}
Denoting the vector with coefficients $\sqrt{s_j} \tr(X_j)$ by $e$, the vector
with coefficients $\sqrt{s_j} \tr(Y_j)$ by $f$ (both of which are real by Hermiticity of $X_i,Y_i$), and the unit vector
with constant coefficients $1/d$ by $g$, we see that we must find $T$ such that:
\begin{eqnarray}
Te = Tf = g.
\end{eqnarray}
Hence $e,f$ must be identical unit vectors if a solution is to exist, and moreover an infinite number of $T$ can be found in such cases. So we simply need to work out when $e,f$ are identical unit vectors. From the fact that:
\begin{eqnarray}
\tr(\rho_{AB}) = 1 = \sum_i s_i \tr(X_i) \tr(Y_i) = e.f \leq \|e\|_2 \|f\|_2,
\end{eqnarray}
and using Cauchy-Schwarz we see that $e,f$ are identical unit vectors iff $\tr(X_i)=\tr(Y_i)$.
\item[B:] Let us represent a quantum state $\sigma$ by $\sigma = \sum_k x_k X_k$, so that the square 2-norm of $\sigma$ is given by $\|\sigma\|^2_2 = \sum_k x^2_k \leq 1$. ${\cal{E}}^{-1}(\sigma)$ is given by:
\begin{equation}
{\cal{E}}^{-1}(\sigma) = \sum_k {1 \over d \sqrt{s_k}} C_k \tr(X^{\dag}_k \sigma).
\end{equation}
A similar equation holds for ${\cal{F}}$. The square of the 2-norm of this expression is given by:
\begin{equation}
\|{\cal{E}}^{-1}(\sigma)\|^2_2 = \sum_k {|x_k|^2 \over d s_k} \leq {1 \over d \min_k s_k},
\end{equation}
so if we have:
\begin{equation}
1/(d \min_k s_k) < d \Rightarrow \min_k s_k > 1/d^2,
\end{equation}
then condition (B) holds. In the case of the maximally entangled state $\min_k s_k = 1/d$, so by continuity there is a region of states around the maximally entangled state that satisfy this.
\end{itemize}


\begin{thebibliography}{70}
\bibitem{strictproof} For convenience we reproduce the following standard argument for showing that the triangle inequality is strict for the norms that we consider:
\begin{eqnarray}
\|X + Y\|^2_{\Lambda} &=& \|X\|^2_{\Lambda} + \|Y\|^2_{\Lambda} + \tr(\Lambda(X)^{\dag}\Lambda(Y)) +  \tr(\Lambda(X)\Lambda(Y)^{\dag}), \nonumber \\
&=& \|X\|^2_{\Lambda} + \|Y\|^2_{\Lambda} + 2\textrm{Re}\{\tr(\Lambda(X)^{\dag}\Lambda(Y))\},\nonumber \\
&\leq & \|X\|^2_{\Lambda} + \|Y\|^2_{\Lambda} +  2 \|X\|_{\Lambda} \|Y\|_{\Lambda}, \nonumber \\
& \leq & (\|X\|_{\Lambda} + \|Y\|_{\Lambda})^2, \nonumber \\
\Rightarrow \|X + Y\|_{\Lambda} & \leq & \|X\|_{\Lambda} + \|Y\|_{\Lambda}.
\end{eqnarray}
The third line follows from the Cauchy-Schwarz inequality, which is strict unless $\Lambda(X) = c \Lambda(Y)$ for some positive number $c$. As $\Lambda$ is taken to be invertible this is equivalent to saying that the inequalities are strict unless $X= c Y$.
\end{thebibliography}
\end{document}